# A Neptune-sized transiting planet closely orbiting a 5–10 million year old star

Trevor J. David[1], Lynne A. Hillenbrand[1], Erik A. Petigura[2], John M. Carpenter[3], Ian J. M. Crossfield[4], Sasha Hinkley[5], David R. Ciardi[6], Andrew W. Howard[7], Howard T. Isaacson[8], Ann Marie Cody[9], Joshua E. Schlieder[9], Charles A. Beichman[6], Scott A. Barenfeld[1]

[1]*Cahill Center for Astronomy and Astrophysics, California Institute of Technology, Pasadena, California 91125, USA*

[2]*Division of Geological and Planetary Sciences, California Institute of Technology, Pasadena, California 91125, USA*

[3]*Joint ALMA Observatory, Av. Alonso de Córdova 3107, Vitacura, Santiago, Chile*

[4]*Lunar & Planetary Laboratory, University of Arizona, Tucson, AZ 85721, USA*

[5]*University of Exeter, Physics Department, Stocker Road, Exeter EX4 4QL, UK*

[6]*NASA Exoplanet Science Institute, California Institute of Technology, Pasadena, California 91125, USA*

[7]*Institute for Astronomy, University of Hawai'i at Mānoa, Honolulu, HI 96822, USA*

[8]*Department of Astronomy, University of California, Berkeley CA 94720, USA*

[9]*NASA Ames Research Center, Mountain View, CA 94035, USA*

**Theories of the formation and early evolution of planetary systems postulate that planets are born in circumstellar disks, and undergo radial migration during and after dissipation of the dust and gas disk from which they formed[1,2]. The precise ages of meteorites indicate that planetesimals – the building blocks of planets – are produced within the first million years of a star's life[3]. A prominent question is: how early can one find fully formed planets like those frequently detected on short orbital periods around mature stars? Some theories suggest the *in situ* formation of planets close to their host stars is unlikely and the existence of such planets is evidence for large scale migration[4,5]. Other theories posit that planet assembly at small orbital separations may be common[6–8]. Here we report on a newly-born, transiting planet orbiting its star every 5.4 days. The planet is 50% larger than Neptune, and its mass is less than 3.6 times Jupiter (at 99.7% confidence), with a true mass likely to be within a factor of several of Neptune's. The 5–10 million year old star has a tenuous dust disk extending in to about 2 times the Earth-Sun separation, in addition to the large planet located at less than 1/20 the Earth-Sun separation.**

USco 161014.75-191909.3, hereafter K2-33, is a several million year old M-type star that was observed by NASA's *Kepler Space Telescope* during Campaign 2 of the *K2* mission. The star was identified as one of more than 200 candidate planet hosts in a systematic search for transits in *K2* data[9]. As part of our ongoing study of the pre-main sequence population of Upper Scorpius observed by *K2*, we independently verified and analyzed the planetary transit signal. We acquired radial velocity (RV) and high spatial resolution observations at the W. M. Keck Observatory to confirm the planet, hereafter K2-33b, and to measure its size and mass.

Within the 77.5 day photometric time series of K2-33 ($Kp$ = 14.3 mag), there are periodic dimmings of 0.23% lasting 4.2 hours and occurring every 5.4 d (Fig. 1). The ensemble of transits are detected at a combined signal-to-noise ratio of $\approx$ 32. During the *K2* observations, cool, dark regions on the stellar surface (starspots) rotated in and out of view, producing semi-sinusoidal brightness variations of ~3% peak-to-trough amplitude with a 6.3 $\pm$ 0.2 d periodicity (Extended Data Fig. 1). We removed the starspot variability prior to modeling the transit events. We fit the transit profiles using established methods[10], measuring the planet's size relative to its host star and its orbital geometry (Table 1).

K2-33 is an established member of the Upper Scorpius OB association[11,12], the nearest site of recent massive star formation ($d$ = 145 $\pm$ 20 pc). Approximately 20% of low mass stars in Upper Scorpius host protoplanetary disks[13], indicating that planet formation is ongoing in the region but in an advanced stage or completed for the majority of stars. The age of the stellar association is 5–10 million years (Myr), as assessed from kinematic, Hertzsprung-Russell (H-R) diagram, and eclipsing binary analyses. The youth of K2-33 itself is established based on the spectroscopic indicators of enhanced hydrogen emission and lithium absorption[11,12], which we confirm from Keck spectra (Table 1). Furthermore, the stellar rotation rate we measure via broadening of absorption lines in the spectra and via the starspot period (Table 1), is rapid relative to field-age stars of similar mass[14]. We determined the star's systemic RV (Table 1) as consistent with the mean value for Upper Scorpius members[15]. Previously, proper motions were used to assess the probability of membership in the association at 99.9%[16]. Finally, the positions of the star in the H-R and temperature-density diagrams (Extended Data Fig. 2) are consistent with the sequence of low mass members of Upper Scorpius[12].

The inferred planet size and mass depend directly upon the host star size and mass. We evaluated the effective temperature and luminosity from our newly determined spectral type (Table 1), extinction-corrected catalog near-infrared photometry, and empirical pre-main sequence calibrations[17,18]. With the temperature and luminosity, we derive a stellar radius from the Stefan-Boltzmann law of $R_* = 1.1 \pm 0.1\ R_\odot$. The radius uncertainty is calculated accounting for recommended errors in temperature[17], photometric errors, and assuming an association depth comparable to its width on the sky. Combining the stellar radius with the planet-to-star radius ratio determined from the *K2* light curve fit, we infer a planetary size for the companion $R_P = 5.8 \pm 0.6\ R_\oplus$, or about 50% larger than Neptune.

We estimate a stellar mass of $M_* = 0.31 \pm 0.05\ M_\odot$ from interpolation among pre-main sequence stellar evolution models[19], consistent with a previously reported value[20]. The mass uncertainty assumes normal error distributions in temperature and luminosity. As there is no evidence for RV variations among four high dispersion Keck spectra (Extended Data Table 1), the planet mass is constrained from the maximum amplitude Keplerian curve that is consistent within the errors with all RV measurements (Extended Data Fig. 3). Given the transit ephemeris and assuming a circular, edge-on orbit, the expected stellar reflex velocity is a sinusoid having a single free parameter: the semi-amplitude. RV semi-amplitudes of $> 900$ m s$^{-1}$ are ruled out at 99.7% confidence, corresponding to a $3\sigma$ upper limit on the mass of K2-33b of 3.6 $M_{\rm Jup}$.

The true mass of K2-33b is likely to be at least an order of magnitude smaller. There are seven known exoplanets of similar size ($R_P$ = 4.8–6.6 $R_\oplus$) with densities measured to 50% or better. These planets have masses ranging from 6.3±0.8 $M_\oplus$ (Kepler-87c) to 69±11 $M_\oplus$ (CoRoT-8b) due to varying core masses. Thus, plausible masses of K2-33b range from ≈6 to 70 $M_\oplus$, corresponding to RV semi-amplitudes of 5–56 m s$^{-1}$. An even lower mass may be implied if the young planet is still undergoing Kelvin-Helmholtz radial contraction.

A semi-major axis of 0.04 AU ($\sim$ 8 $R_*$) is measured for K2-33 from the orbital period and Kepler's third law, adopting $M_*$ (Table 1). The orbit is near the silicate dust sublimation radius, as well as the co-rotation radius, where some protoplanetary disk theories predict a magnetospheric cavity extending to the stellar surface[21]. At this separation, the blackbody equilibrium temperature of the planet is 850 K.

We have interpreted the observed transit as a single planet orbiting K2-33. Other interpretations involving eclipsing binaries (EBs) residing within the $K2$ aperture (Extended Data Fig. 4) would be diluted by the light of K2-33 and could potentially mimic the observed transit. Such a putative EB could be associated with (i.e. gravitationally bound to) K2-33, or unassociated but aligned by chance. Given constraints from our suite of follow-up observations, we show that the chance an EB produces the observed transit is vanishingly small.

We first consider unassociated EBs, which we characterize by their sky-projected separation from K2-33 and their brightness relative to K2-33 in the Kepler bandpass, $\Delta Kp$. Eclipse depths may not exceed 100%; thus EBs with $\Delta Kp > 6.6$ mag cannot account for the 0.23% observed transit depth. We do not detect closely-projected companions in seeing-limited and multi-epoch adap-tive optics images (Extended Data Fig. 5), nor in searches for secondary lines in high-resolution optical spectra. These observational constraints, shown in Fig. 2a, eliminate nearly all scenarios involving unassociated EBs. The probability of an EB lurking in the remaining parameter space is $<< 4 \times 10^{-6}$ (see Methods).

We now consider associated EBs in terms of their physical distance, $d$, to K2-33 and $\Delta Kp$. As in the case of unassociated EBs, our imaging and spectroscopic data eliminate the vast majority of associated EB configurations. Additionally, the lack of detectable line-of-sight acceleration over the baseline of the observations rules out associated EBs in a mass-dependent manner (see Methods). The constraints provided by these complementary observations are depicted in Fig. 2b. However, some scenarios having $d$ = 1–3 AU and $\Delta Kp$ = 2–6 mag, cannot be conclusively eliminated. Nearly all of these scenarios involve a planet orbiting either K2-33 or an undetected companion. If orbiting K2-33, the planet radius is at most 7.6% larger (within quoted uncertainties) due to dilution from a companion. If orbiting a stellar or substellar companion to K2-33, the planet radius is at most $\sim$1.8 $R_{\rm Jup}$. Only for $\Delta Kp \geq 6$ mag does the radius of the transiting object corresponds to a mass $\gtrsim 13$ $M_{\rm Jup}$ at 5–10 Myr. However, coeval eclipsing brown dwarfs likely would not produce eclipse depths >50%, and thus contrasts of $\Delta Kp \geq 5.8$ mag need not be considered. Furthermore, the low occurrence of brown dwarfs[22] combined with the lack of observed secondary eclipses make such configurations extremely unlikely. Given the *Kepler* photometry and observational constraints, we quantitatively assessed the overall false positive probability, using an established statistical framework[23] (see Methods). We found scenarios involving a single star and planet are $10^{11}$ times more likely than scenarios involving EBs.

*Spitzer Space Telescope* observations of K2-33 revealed 24 micron emission in excess of the expected stellar photosphere by 50%, indicating the presence of cool circumstellar dust[24]. There is an absence, however, of warm dust close to the star given the lack of similar infrared excess at wavelengths shorter than 16 microns[25]. The spectral energy distribution is best fit by including a dust component at 122 K having an inner edge at 2.0 AU. These data suggest that the inner regions of the previously present protoplanetary disk have cleared. Supporting this inference is the modest H$\alpha$ signature (Table 1), which is consistent with chromospheric emission and indicates the star is not accreting gas.

At the age of K2-33, it is unclear whether the dust structure consists of debris resulting from the collisional grinding of planetesimals, or if it is a remnant of the initial dust and gas-rich disk. One possibility is that the inner disk regions have been opened due to the dynamical effects of one or more planetary mass bodies[26]. Our detection of a short period planet in a transitional disk lends support to this explanation.

A flux upper limit at 880 microns from the Atacama Large Millimeter Array[27] combined with the measured *Spitzer* fluxes yields a constraint on the mass of dust remaining in the disk of less than 0.2 $M_\oplus$. Additionally, CO emission, a tracer of molecular hydrogen, was not detected[27], indicating the primordial gas disk also has largely or entirely dissipated.

The transiting planet around the young star K2-33 provides direct evidence that large planets can be found at small orbital separations shortly after dispersal of the nebular gas. Migration via tidal circularization of an eccentric planet, through e.g. the Kozai-Lidov mechanism, planet-planet scattering, or secular chaos, proceeds over timescales much greater than disk dispersal timescales, and thus cannot explain the planet's current orbit. In situ formation or formation at a larger separation followed by migration within the gas disk are permitted scenarios given current observations.

Interestingly, large planets are rarely found close to mature low mass stars; fewer than 1% of M-dwarfs host Neptune-sized planets with orbital periods <10 days[28], while ∼20% host Earth-sized planets in the same period range[29]. This may be a hint that K2-33b is still contracting, losing atmosphere, or undergoing radial migration. Future observations may test these hypotheses, and potentially reveal where in the protoplanetary disk the planet formed.

**Acknowledgements** We thank the referees for insightful comments that greatly improved this work, K. Batygin, B. Benneke, K. Deck, J. Fuller, S. Metchev, and A. Shporer for discussions, M. Ireland for software used in the aperture masking analysis, and A. Kraus for contributing to the 2011 Keck/NIRC2 data acquisition. T.J.D is supported by an NSF Graduate Research Fellowship under Grant DGE1144469. E.A.P.

is supported through a Hubble Fellowship. This paper includes data collected by the *Kepler* mission, funded by the NASA Science Mission directorate. Some data presented herein were obtained at W.M. Keck Observatory, which is operated as a scientific partnership among the CIT, the Univ. of California and NASA. The authors recognize and acknowledge the significant cultural role and reverence that the summit of Mauna Kea has always had within the indigenous Hawaiian community. We are most fortunate to conduct observations from this mountain.

**Author contributions**   T.J.D. noted the object as a young star, prepared the light curve, validated the transit, and led the overall analysis and paper writing. L.A.H. analyzed the 2015 and 2016 Keck/HIRES spectra, participated in team organization, performed general analysis, and contributed significantly to paper writing. E.A.P. analyzed raw *K2* photometry and validated the transit, provided general guidance on exoplanets and false positives, and contributed significantly to paper writing. J.M.C. was involved in our *K2* proposal that included the object, and was P.I. on the 2011 Keck/NIRC2, 2015 Keck/HIRES, and ALMA observations of the object. I.J.M.C. led the transit fitting and VESPA analysis. A.M.C. analyzed the raw *K2* photometry and validated the transit. A.W.H. and H.T.I. obtained, reduced, and analyzed the 2016 Keck/HIRES spectra. D.C. led the clear aperture adaptive optics contrast curve analysis for the 2016 and 2011 data. C.A.B. wrote the proposal for 2016 Keck/NIRC2 follow-up of *K2* sources and participated in the observations of the object. S.H. analyzed the Keck/NIRC2 aperture masking data and assessed the temperature of the circumstellar dust. J.E.S. provided a rotational velocity analysis and calculated the kinematic distance. S.A.B. took the 2015 Keck/HIRES spectrum and analyzed the ALMA data that is referred to, but published separately.

**Author information**   The authors declare that they have no competing financial interests. Correspondence and requests for materials should be addressed to T.J.D. (email: tjd@astro.caltech.edu).

Table 1: System properties of K2-33

| Parameter | Value | Uncertainty |
|---|---:|---|
| **Stellar properties** | | |
| 2MASS designation | J16101473-1919095 | |
| EPIC designation | 205117205 | |
| Right ascension (J2000.0) | $16^h10^m14.738^s$ | |
| Declination (J2000.0) | -19°19'09.55" | |
| Proper motion R.A., $\mu_\alpha$ (mas/yr) | -9.8 | $\pm 1.7$ |
| Proper motion Dec., $\mu_\delta$ (mas/yr) | -24.2 | $\pm 1.8$ |
| $Kp$ (mag) | 14.3 | |
| Cluster distance, d (pc) | 145 | $\pm 20$ |
| Kinematic distance, $d_{kin}$ (pc) | 139 | $\pm 11$ |
| Spectral type | M3 | $\pm 0.5$ |
| V-band extinction, $A_V$ (mag) | 1.29 | |
| Luminosity, $\log L_*/L_\odot$ (dex) | -0.83 | $\pm 0.07$ |
| Effective temperature, $T_{eff}$ (K) | 3410 | $\pm 75$ |
| Stellar radius, $R_*$ ($R_\odot$) | 1.1 | $\pm 0.1$ |
| Stellar mass, $M_*$ ($M_\odot$) | 0.31 | $\pm 0.05$ |
| Mean stellar density, $\rho_*$ (g/cc) | 0.34 | $\pm 0.12$ |
| Surface gravity, $\log g_*$ (dex) | 3.84 | $\pm 0.16$ |
| Stellar rotation period, $P_{rot}$ (d) | 6.3 | $\pm 0.2$ |
| Systemic radial velocity, $\gamma$ (km s$^{-1}$) | -6.6 | $\pm 0.1$ |
| Projected rotational velocity, $v\sin i$ (km s$^{-1}$) | 5–9 | |
| H$\alpha$ equivalent width (Å) | -1.3 | $\pm 0.1$ |
| H$\beta$ equivalent width (Å) | -1.05 | $\pm 0.05$ |
| LiI 6708Å equivalent width (Å) | 0.60 | $\pm 0.05$ |
| **Light curve modeling parameters** | | |
| Orbital period (days) | 5.42513 | $^{+0.00028}_{-0.00029}$ |
| Time of mid-transit, $t_0$ (BJD$_{TDB}$; days) | 2,456,936.6665 | $^{+0.0012}_{-0.0012}$ |
| Transit duration, $T_{14}$ (hours) | 4.22 | $^{+0.15}_{-0.10}$ |
| Planet-to-star radius ratio, $R_p/R_*$ | 0.0476 | $^{+0.0035}_{-0.0017}$ |
| Scaled semi-major axis, $R_*/a$ | 0.109 | $^{+0.033}_{-0.012}$ |
| Impact parameter, b | 0.49 | $^{+0.26}_{-0.33}$ |
| Inclination, i (deg) | 86.9 | $^{+2.2}_{-3.1}$ |
| Mean stellar density, $\rho_{*,circ}$ (g/cc) | 0.49 | $^{+0.21}_{-0.27}$ |
| Linear limb darkening coefficient, $u$ | 0.603 | $^{+0.052}_{-0.053}$ |
| **Planet properties** | | |
| Planet radius, $R_p$ ($R_\oplus$) | 5.76 | $^{+0.62}_{-0.58}$ |
| Planet mass, $M_p$ ($M_{Jup}$) | <3.6 | |
| Semi-major axis, $a$ (AU) | 0.0409 | $^{+0.0021}_{-0.0023}$ |
| Blackbody equilibrium temperature, $T_{eq}$ (K) | 850 | $\pm 50$ |

Right ascension and declination originate from 2MASS, proper motions from UCAC4, and Kepler magnitude from the Ecliptic Plane Input Catalog. The mean cluster distance[30] is assumed, with uncertainty equal to the presumed cluster depth[12]. Quoted transit parameters and uncertainties are medians and 15.87%, 84.13% percentiles of the posterior distributions.

**Figure 1. Light curve of K2-33.** Light curve of K2-33. **a-d**, *K2* photometry in twenty day segments. K2-33 varies in brightness by ∼3% every 6.3 days due to rotation of its spotted surface. The planet K2-33b transits its star every 5.4 days (red ticks). Another potential transit (blue tick), distinct from the K2-33b transits, is possibly due to a second planet with orbital period > 77.5 days. **e**, Stellar variability was removed prior to fitting transits. The data gap is due to excluded observations where the variability fit inadequately captured a systematic artifact in the light curve. **f**, *K2* photometry folded on the planet's orbital period with a transit model fit (red curve).

**Figure 2. Constraints on astrophysical false positive scenarios.** To confirm the planetary nature of K2-33b, we considered and eliminated nearly all false positive scenarios involving eclipsing binaries (EBs). **a**, The domain of sky-projected separation and contrast of a putative unassociated EB, aligned with K2-33 by chance. The blue hatched region shows EBs eliminated using multi-epoch AO imaging, which leverages stellar proper motion to provide sensitivity at all separations within 5". Green and purple regions represent constraints from optical spectra and seeing-limited imaging, respectively. Finally, EBs in the gray region cannot account for the observed transit depth and are eliminated. **b**, Limits on EBs associated with (i.e. gravitationally bound to) K2-33. Constraints from imaging and spectroscopy are shown as a function of physical separation. The lack of detectable stellar acceleration provides an additional diagonal constraint at top left.

**Methods**

**Stellar Membership and Properties.** The partial kinematics of K2-33 (Table 1) can be combined with the galactic velocity of the Upper Scorpius subgroup to estimate distance and predict radial velocity (RV). We calculated these parameters following established methods[31], adopting the mean UVW galactic velocities of the subgroup[32], and estimated uncertainties using Monte Carlo sampling. We find $d_{\rm kin} = 139 \pm 11$ pc, consistent with the mean subgroup distance $d = 145 \pm 2$ pc[30], and RV $= -7.3 \pm 0.5$ km s$^{-1}$, within $2\sigma$ of the systemic RV we measure from multiple Keck/HIRES spectra (Extended Data Table 1).

We determined spectral type from the high dispersion spectra and adopted an empirical spectral type to temperature conversion calibrated for young stars[17,18] to estimate effective temperature (Table 1). From 2MASS photometry and the appropriate intrinsic $J - H$ color for the spectral type[18], we calculated the $J - H$ color excess, $E(J - H) = 0.10$ mag. Assuming an extinction law we found visual and $J$-band extinctions of $A_V$ = 1.29 mag and $A_J$ = 0.30 mag, respectively. After correcting for extinction, we used the appropriate empirical $J$-band bolometric correction for the spectral type[17,18] and a distribution of distances to calculate luminosity. While the mean association distance is known precisely, its large sky-projected area (about 150 deg$^2$) suggests the association depth is significant. Results from a secular parallax study indicate an association distance spread <50 pc, with consideration of the angular diameter corresponding to 35 pc spread [12,33] at the mean distance. In calculating luminosity, we conservatively considered a uniform distribution of distances in a cubic volume 40 pc on each side, centered on the mean association distance of 145 pc[30]. Uncertainty is calculated from Monte Carlo sampling, accounting for photometric errors, recommended errors in temperature[17], and distance uncertainties. We propagated luminosity and temperature errors through in calculating the radius uncertainties using the Stefan-Boltzmann law.

Detailed modeling of the transit profile provides a constraint on mean stellar density, assuming a circular orbit[34]. We found a mean stellar density of $\rho_{*,\mathrm{circ}}$ = 0.49 g cc$^{-1}$, consistent with the value implied by our adopted $M_*$ and $R_*$ (Table 1). From the posterior distribution of mean stellar densities and a normal distribution in effective temperature, we interpolated between pre-main sequence models[19] to determine a stellar mass and age of $M_* = 0.30 \pm 0.04 M_\odot$ and $\tau = 5.1^{+1.5}_{-2.6}$ Myr, respectively (Extended Data Fig. 2). However, we conservatively adopt the ensemble age of the association, as we consider it more robust than the age of an individual star.

**Stellar Rotation and Independent Assessment of the Stellar Radius.** We measured the stellar rotation period as $P_{\mathrm{rot}} = 6.3 \pm 0.2$ d from a Lomb-Scargle periodogram[35,36] of the light curve. Uncertainty was determined from the half width at half maximum of a Gaussian fit to the periodogram peak. Extended Data Fig. 1 depicts the light curve folded on the rotational period.

The stellar rotation speed as projected along the line of sight, $v \sin i_*$, was estimated from the spectra by artificially broadening an absorption spectrum of a slowly rotating stellar template of similar temperature to K2-33, acquired using the same spectrograph and set-up. The range of plausible rotational velocities is constrained through minimization of the residuals between the broadened template and the observed spectrum. We find a most likely projected rotation velocity $v \sin i_* \approx$ 5–9 km s$^{-1}$. Combined with the rotation period measured from the light curve, we used the projected rotational velocity to determine an independent estimate of the stellar radius modulated by the sine of the stellar inclination of $R_* \sin i_* = v \sin i_* \cdot P_{\mathrm{rot}}/2\pi = 0.85 \pm 0.25\ R_\odot$, where we quote a 95% confidence interval assuming a uniform distribution in $v \sin i_*$.

Our value for $R_* \sin i_*$ is consistent with the Stefan-Boltzmann radius within 2-$\sigma$. Two effects could bias $R_* \sin i_*$ away from the true value of $R_*$: (1) the surface features dominating the rotational modulation of the light curve may be confined to a range of stellar latitudes that may not reflect the same velocity field encoded in the rotational broadening of spectral lines, and (2) the star may have an inclination resulting in $\sin i_*$ significantly $< 1$. If this is the case, the orbit of K2-33b is misaligned with the spin of its host star. While $R_* \sin i_*$ provides a valuable consistency check on stellar radius, we use the Stefan-Boltzmann radius given its insensitivity to the unknown stellar inclination.

*K2* **Time Series Photometry Treatment.** The *K2* mission[37] observes fields along the ecliptic plane for approximately 75 days at a time. *K2* photometry possesses percent-level systematic signatures from pointing drift and intra-pixel detector sensitivity variations that must be corrected in order to detect sub-percent planet transits. We acquired such a corrected light curve from the Exoplanet Follow-up Observing Program public website[9], derived using a 12" × 16" rectangular aperture (Extended Data Fig. 4).

Before modeling the transit profile, we removed the spot modulation pattern using a cubic basis spline fit with knots spaced by 12 long-cadence measurements, employing iterative rejection of outliers[38]. We verified that no in-transit observations were included in the spline fit by phasing the flattened light curve on the orbital period and inspecting the points excluded from the fit. An artificial gap in the systematics corrected light curve from the ExoFOP page was not adequately captured by the spline fit and, consequently, we excluded from further analysis those data with BJD values in the range 2,456,936 to 2,456,938 resulting in the loss of a single transit. We assigned a constant observational uncertainty for each *K2* measurement, determined from the standard deviation in the out-of-transit light curve (here defined as observations more than 12 hours from either side of the transit centers).

**Transit Model Fitting Analysis.** We employed previously established methodology[10] for fitting transit models to the light curve. The approach uses the BATMAN software[39], based on the Mandel & Agol analytic light curve formalism[40], to generate model transit profiles. Transit models were numerically integrated to match the ≈30 minute cadence of *K2* observations. We assumed a linear limb-darkening law for the star, imposing a Gaussian prior on the linear limb-darkening coefficient, $u$, based on tabulated values[41] appropriate for the effective temperature and surface gravity of K2-33. We also allowed for dilution by light from a second star in the fitting. In post-processing, we selected only those samples corresponding to dilution levels consistent with our companion exclusion analysis (Fig. 2).

The directly fitted transit parameters are: scaled semi-major axis ($a/R_*$), planet-to-star radius ratio ($R_p/R_*$), orbital period ($P$), time of mid-transit ($t_0$), and inclination ($i$). The multi-dimensional transit parameter space was explored using an affine invariant implementation of the Markov chain Monte Carlo (MCMC) algorithm[42] to find the best-fit model and determine parameter uncertainties. Each observation was weighted equally, resulting in a best-fit likelihood of $-\frac{1}{2}\chi^2$. We followed established methods[10] for MCMC chain initialization, burn-in treatment, rescaling of data weights, and convergence testing. Table 1 quotes median transit model values, with uncertainties determined from the 15.87% and 84.13% percentiles of the parameter posterior distributions.

The scatter in transit is ∼15% larger than the scatter in equal duration intervals before and after transit; one possible explanation may be spot-crossing events, given that the star is expected to have a high spot-covering fraction[43], supported by the modulation in the unflattened light curve.

Prior to adopting the publicly available light curve[9], we independently extracted a light curve from *K2* target pixel files using a different photometric pipeline[10,44]. Performing the same modeling described above on this second light curve produced consistent results.

**High Spectral Resolution Observations and Radial Velocities.** We used the HIgh REsolution Spectrograph[45] (HIRES) on the 10-meter Keck-1 telescope to measure the radial velocity of K2-33 relative to the Solar System barycenter (Extended Data Table 1) in order to confirm cluster membership, and to constrain the mass of K2-33b. The resolution of these spectra are $R \approx 50,000$ between ∼3600–8000 Å for the 2016 epochs and $R \approx 36,000$ between ∼4800–9200 Å for the 2015 epoch.

For the first epoch, velocity was derived by cross-correlating the spectrum with radial velocity standards[46] observed using HIRES in the same spectrograph configuration. Uncertainty is quantified from the dispersion among measurements relative to different standards, and over many different spectral orders. For the three epochs in 2016, systemic radial velocity was measured using the telluric A and B absorption bands as a fiducial wavelength reference. Assessing all measurements, the star's systemic radial velocity is -6.6 ± 0.1 km s$^{-1}$, where we quote a weighted average and standard deviation.

**Limits on Companions from the Spectroscopic Data.** We searched for and excluded distant gravitationally bound companions to K2-33 by looking for trends in the 265-day RV time series. A $3\sigma$ upper limit to any possible acceleration is 2.6 km s$^{-1}$ yr$^{-1}$. Following established methods[47], the limiting minimum mass detectable from the RV measurements as a function of orbital separation rules out additional companions of $\geq 0.14$ $M_\odot$ at 3 AU and $\geq 0.39$ $M_\odot$ at 5 AU.

We also searched for secondary spectral lines[48] that would arise from a companion projected within 0".4 of the primary. No stars as faint as 3% the brightness of the primary were detected, though we are blind to companion stars with a small ($<15$ km s$^{-1}$) radial velocity relative to the primary since the spectral lines of the two stars would not be distinguished.

**High-resolution Imaging.** Using adaptive optics at the Keck-2 telescope on UT 2011 May 15, we obtained ten 9 second exposures of K2-33. A second set of Keck/NIRC2 images was acquired on UT 2016 February 17 in a three-point dither pattern with 10 second integrations per dither position, repeated three times. For both epochs, the narrow-camera optics were used resulting in a pixel scale of 9.942 mas pixel$^{-1}$; the final coadded images have resolution 0.07" (full width at half maximum). The observations at the two epochs have the same total integration time, so the final images have comparable depth.

To estimate sensitivity to point sources as a function of radial distance from the star, median flux levels and root-mean-square dispersions ($\sigma$) were calculated in incremental annuli centered on the source. An image was constructed with these characteristics and synthetic sources with full width at half maximum equivalent to that of K2-33 were injected into the image at varying positions and brightnesses. The synthetic sources were then measured to determine the $5\sigma$ detection limit. Comparing the instrumental magnitude to that of the star produces $\Delta K$ for that annulus.

**Non-redundant Aperture Masking.** K2-33 was observed using an aperture mask on the same 2011 night as the clear aperture images were taken. Aperture masking interferometry[49] uses an opaque mask with clear holes such that the baseline between any two samples a unique spatial frequency in the pupil plane, and achieves angular resolution as good as $\frac{1}{3}\lambda/D$ compared to 1.2 $\lambda/D$ though at the expense of diminished throughput[50,51].

With the nine-hole mask in the NIRC2 camera, we obtained 40 dithered images using 20 second exposures. Observations of calibrator stars were interspersed with the targets and obtained in an identical manner. In each image, the mask creates a set of 36 overlapping interference fringes on the detector. The bispectrum, the complex triple product of visibilities defined by the three baselines formed from any three subapertures, is then calculated. The phase of this complex quantity is the closure phase, which has the advantage of being largely insensitive to phase delays due to atmospheric effects or residual uncorrected phase aberrations not sensed by the adaptive optics system. We followed established procedures[52,53] for calculating the closure phase for calibrator stars.

**Limits on Companions from the Imaging Data.** Using all three sets of high spatial resolution near-infrared imaging data, we searched for projected companions to K2-33 (either gravitationally bound or forground/background sources). The two epochs of clear aperture data each achieved $\Delta K > 4.5$ mag of contrast beyond 0".13 and $\Delta K > 7.5$ mag beyond 1". In the interim between the observations, K2-33 moved on the sky by 0."1228 ± 0."0085 due to its parallactic and proper motion. The combined set of infrared images thus rules out nearly all unassociated background stars bright enough to produce the transit observed in the optical. The aperture masking observations are sensitive to more closely projected (from 0".02-0".16) companions in the stellar and substellar mass regimes. We use these to rule out potential associated companions larger than 19 Jupiter masses down to orbital separations of 3 AU, and are sensitive to companions with masses as low as 11–12 Jupiter masses from 6–23 AU, where quoted companion masses are model-dependent[19] conservatively assuming an age of 10 Myr. Finally, a prior analysis of near-infrared photometry ruled out associated companions down to masses of 5-6% of the primary mass at separations of 1.7"-27.5", or approximately 250-4000 AU[54].

The imaging limits constrain the brightnesses of any putative companions in the near-infrared $K$-band. To approximately convert these contrast limits to constraints in the optical *Kepler* bandpass, $Kp$, we employed a combination of theoretical evolutionary models and empirical color-color relations. The TRILEGAL simulation discussed below predicts that the mean potential contaminant toward K2-33 would have infrared color $J - K_s = 0.57$ mag, corresponding to a K-type dwarf. From an empirical optical-infrared $Kp - K_s$ conversion[55], we conclude that a typical contaminating source would have color $Kp - K_s = 2.00$ mag. We thus gain an additional 2 mag of contrast when converting the NIRC2 contrast curve to the corresponding limits in the Kepler bandpass when considering unassociated companions.

For associated companions, the imaging constraints natively derived in the near-infrared were converted to optical limits using pre-main sequence evolutionary models[19]. Putative companion masses can be paired with our assumed primary mass to yield predicted $R$-band contrasts at 5–10 Myr, where $R$-band serves as a proxy for the *Kepler* bandpass. Notably, for the clear aperture AO region of Fig. 2b, the achieved contrast beyond 30 AU is better than represented in the figure due to limitations of the models which do not extend below 10 $M_{\rm Jupiter}$ (corresponding to $\Delta\,Kp \gtrsim 9.5$ mag).

We rule out nearly all stellar and brown dwarf mass companions to K2-33, with the exception of a narrow swath of parameter space corresponding to ultra low mass stars or brown dwarfs separated by 1–3 AU from the primary (Fig. 2). The physically permitted (as opposed to unexcluded) separation range of any hypothetical diluting companion is even smaller, considering the inferred inner edge of the disk at 2 AU.

**Galactic Structure Model and Intracluster Contamination.** In addition to searching directly for sources that could contaminate the *K2* light curve, we estimated the expected surface density of such sources as a function of magnitude using the TRILEGAL v1.6[56,57] model of the Milky Way Galaxy. Notably, TRILEGAL does not include the local extinction due to gas and dust associated with Upper Scorpius itself, and therefore produces an upper limit to the field star source density. We simulated a 1 deg$^2$ field and scaled the resulting numbers **first** to the 10 $\times$ 10 arcsec$^2$ field of view of the Keck/NIRC2 images and the 12 $\times$ 16 arcsec$^2$ *K2* photometric aperture, and then to the unexcluded regions of parameter space in Fig. 2a. We found the TRILEGAL prediction for the expected surface density of sources to the $K_s < 18.0$ mag limit (5$\sigma$) of the NIRC2 data, and translated it to <0.15 sources per NIRC2 field, consistent with our detection of none. Within the *K2* photometry aperture, <0.3 unassociated sources are expected to the same magnitude limit, corresponding to a mean optical brightness $Kp$ = 18.6 mag.

By considering the surface density of sources expected from integrating essentially all the way through the galaxy, a maximum of 3 sources are expected in the *K2* aperture, having mean optical brightness $Kp$ = 23.1 mag which is too faint to explain the transit depth. Indeed, we have ruled out nearly all companions with projected separations larger than 0.04", as well as effectively all of those inside of 0.04"; fewer than $4 \times 10^{-6}$ sources are expected in the remaining unexcluded parameter space of Fig. 2a. An even smaller number of sources are expected to be EBs. Therefore, in addition to not detecting any contaminants in the high spatial resolution imaging and spectroscopic data, we argue that essentially none are expected.

A similar source density argument can be used to constrain the probability of contaminants with nearly identical proper motions to K2-33, likely association members that are foreground or background to K2-33. The multi-epoch AO imaging cannot rule out closely-projected sources within 0.02" (the inner limit probed by the aperture masking observations) that are also co-moving with K2-33 between 2011–2016. From the observed Upper Scorpius mass function and width of the association on the sky[12], we estimated a source density of $\sim$ 16 members deg$^{-2}$. Thus, fewer than $2 \times 10^{-9}$ co-moving contaminants are expected within 0.02" of K2-33.

**False Positive Probability Analysis.** Eclipsing stellar binaries (EBs) when diluted by the light of a third star can produce light curves that masquarade as a planetary transit. These false positives come in three broad classes: (1) undiluted EBs, (2) background (and foreground) EBs where the eclipses are diluted by the target star (3) bound EBs in hierarchical triple systems.

We used the VESPA program[23] to compare the likelihood of each binary scenario against the planetary interpretation. As input for the calculation, we provide the *K2* light curve, the stellar parameters, and to be as conservative as possible, we adopt our least stringent imaging constraints: the 2011 NIRC2 clear contrast curve, and the aperture masking limits. Even in this minimally constraining scenario, we find a false positive probability of $< 1 \times 10^{-11}$ from VESPA, as expected given our exclusion in Fig. 2 of essentially all the modelled scenarios. Notably, however, VESPA does not account for substellar objects, pre-main sequence evolution, or extinction, nor the unknown prior probability of planets around 5–10 Myr stars.

**Implications of Hierarchical Triple Scenarios.** The conjectured hierarchical triple configuration is argued elsewhere as unlikely based on population statistics; however, we must still consider the possibility that K2-33b has a larger radius due to dilution of the transit depth by a luminous companion to K2-33. We first consider the case in which the planet orbits K2-33, but the transit depth is diluted by an undetected secondary. In this scenario, the ratio of the true planet radius to the observed planet radius (i.e. not accounting for dilution) is $R_{P,\text{true}}/R_{P,\text{obs}} = \sqrt{1 + F_{\text{sec}}/F_{\text{pri}}}$, where $F_{\text{sec}}/F_{\text{pri}}$ is the optical flux ratio between the secondary and primary. For $\Delta Kp$ between 2–6 magnitudes, the true planet radius is at most 7.6% larger, within our quoted uncertainties.

Now, we consider the case in which the planet orbits an undetected companion to K2-33. In this scenario, the planet radius is $R_P = R_{\text{sec}}\sqrt{\delta_0(1 + F_{\text{pri}}/F_{\text{sec}})}$, where $R_{\text{sec}}$ is the secondary radius, and $\delta_0$ is the observed transit depth. Using $\Delta R$ as a proxy for $\Delta Kp$, we found the secondary radius implied by the optical brightness decrement using evolutionary models valid for 5–10 Myr[19]. For $\Delta Kp$ between 2–6 magnitudes, we found the implied planet radius is in the range 0.56–1.85 $R_{\rm Jup}$. At these ages, such radii correspond to planet masses of $\lesssim$13 $M_{\rm Jup}$. Only for $\Delta Kp \gtrsim 6$ mag does the implied mass exceed the nominal brown dwarf minimum mass. However, coeval eclipsing brown dwarfs are unlikely to produce eclipse depths >50% (corresponding to contrasts of $\Delta Kp \gtrsim$ 5.8 mag). Furthermore, we argue that scenarios involving eclipsing brown dwarfs 1-3 AU from K2-33 are extremely unlikely, due to several compounding reasons: (1) the observed frequency of brown dwarf companions to M-dwarfs is a few percent[22,58]; (2) in the restricted domain of $a = 1 - 3$ AU, the frequency is lower still; and (3) the frequency of eclipsing brown dwarf pairs so similar in size and temperature that primary and secondary eclipse are indistinguishable is smaller still. Finally, the tenuous dust disk with inner edge at 2 AU is further evidence in favor of the single star scenario.

**Cluster Age.** The age of Upper Scorpius is constrained to be between 5–10 Myr from a variety of considerations. Absence of dense molecular gas or protostars in Upper Scorpius implies that star formation has ceased in the region[59], while presence of protoplanetary disks around a significant fraction of members indicates that planet formation is ongoing[13]. However, the precise age of the association is currently debated. An early kinematic analysis, in which the motions of high mass members were traced back in time to the point of closest proximity to one another, suggested an age of 5 Myr[60]. The first H-R diagram analysis of the full stellar population, spanning from the highest to lowest masses, also determined an age of 5 Myr without significant dispersion[12]. Most subsequent age determinations using theoretical evolutionary models in the H-R diagram arrived at the same consistent age of 5-6 Myr for massive main sequence turnoff stars[61,62], low mass stars[17,63–65], as well as substellar mass objects[66].

However, the association age has also been determined to be closer to 10 Myr from analyses of low mass members[67,68] and the intermediate mass pre-main sequence and main sequence population, main sequence turn-off stars, and the supergiant Antares[69]. Emerging evidence from double-lined eclipsing binaries also supports an age in the range of 7-11 Myr[38,70,71], as does an updated kinematic analysis[69]. Despite the lack of consensus on the precise age of Upper Scorpius, the full error-inclusive range of estimates in the literature (3-13 Myr) place the association at a critically important stage in the planet formation process – when most primordial disks have dispersed[72].

**K2-33b in the Context of Other Claimed Young Planets.** While several secure short-period planets have been found in orbit around stars in benchmark open clusters including the 600-800 million year old Hyades[73,74] and Praesepe[75,76], the evidence for planets at younger ages is mixed. Direct imaging has revealed 5-10 $M_{\rm Jupiter}$ "planetary mass companions" located at large semi-major axes from several stars having ages a few tens to a few hundred million years. Additionally, there are strong indications of ongoing planet formation in many 1–3 million year old circumstellar disks based on observed radial structure of dust.

However, there are no confirmed planets in the Jupiter or sub-Jupiter mass range with ages less than those corresponding to the late-heavy bombardment era in our own solar system. Both TW Hya and PTFO 8-8695 have been claimed to host hot Jupiter candidates detected via radial velocity or transit methods, but neither object has stood up to scrutiny[77–79]. K2-33b at 5–10 Myr of age, by contrast, is a secure transiting planet. It is slightly larger than Neptune and probably Neptune-mass within a factor of several.

**Code availability.** We have opted not to make available our custom codes employed in the various elements of our data reduction and analysis, as they are either not of general purpose use, lack sufficient documentation, or they are already publicly available. The results of the computer processing are provided as Source Data. The raw data is publicly available, either currently (*K2*) or after a limited proprietary period (Keck, ALMA).

**Extended Data Table 1. Keck/HIRES radial velocities for K2-33.** Line-of-sight velocities and $1\sigma$ uncertainties (standard deviations) with respect to Solar System barycenter.

**Extended Data Figure 1.** *K2* **light curve for K2-33 phased on the stellar rotation period of 6.3 d.** Semi-sinusoidal brightness variations due to rotational modulation of starspots. Point color indicates relative time of observation. Brightness is lowest when the most heavily spotted hemisphere of the stellar surface is along the line of sight. The shape and evolution of the variability pattern depends on the number, geometry, distribution, and lifetime of spots, along with any latitudinal gradient in the rotational speed (differential rotation). The transits of K2-33b are visible by eye in this figure and are too narrow in rotational phase to be attributed to any feature on or near the stellar surface.

**Extended Data Figure 2. Model-dependent age of K2-33. a**, Solid lines show mean stellar density as a function of effective temperature for pre-main sequence stars having different ages, according to theoretical models[19]. Grey points represent plausible combinations of density and temperature for K2-33 as determined by light curve fits and stellar spectroscopy. **b**, Distribution of implied stellar age based on temperature, density, and pre-main sequence models. The implied age of 2–7 Myr is consistent with our adopted age of 5–10 Myr, derived independently. Dark and light grey shaded regions indicate 68% and 95% confidence intervals, respectively.

**Extended Data Figure 3. Apparent radial velocity variations of K2-33.** Line-of-sight velocities and $1\sigma$ uncertainties (standard deviations) with respect to Solar System barycenter from Keck/HIRES are indicated. Radial velocities (RVs) are mean-subtracted, and the abscissa shows the orbital phase of K2-33b measured from *K2* photometry (mid-transit occurs at zero orbital phase). We rule out RV variations larger than 300 m s$^{-1}$ at 68.3% confidence, corresponding to a 1.2 $M_{\rm Jupiter}$ planet mass. Curves show expected RV variations for planets having circular orbits and different masses. RVs due to a 1.0 Jupiter mass planet (blue) are consistent with our observations, while a 4.0 Jupiter-mass planet is ruled out at high confidence.

**Extended Data Figure 4. Images of K2-33.** **a**, *K2* target pixel file (TPF). **b**, Sloan Digital Sky Survey (SDSS) optical image. **c**, Keck/NIRC2 $K$-band image. Extents of the *K2* TPF, *K2* photometric aperture, and NIRC2 image are shown respectively with black, green, and purple boundaries. In each image, north is up and east is left. Three other sources identified by SDSS reside within the *K2* photometric aperture, one of which is a galaxy. All are 7.3–10.1 magnitudes fainter than K2-33 in the SDSS $r$-filter and below the detection limit of the NIRC2 images, and thus too faint to produce the observed transits.

**Extended Data Figure 5. Sensitivity to non-comoving sources in the vicinity of K2-33.** The blue X marks the star's position in 2011. Between 2011 and 2016, the star moved by 0."1228 ± 0."0085 (red X) due to proper motion. Contours show $K$-band sensitivity to non-comoving stars from adaptive optics imaging from both epochs. The 2011 dataset included non-redundant aperture masking, and provided tighter constraints. The combined sensitivity to non-comoving objects is the maximum contrast achieved for either dataset. Due to stellar proper motion, we achieved $K$-band contrasts of >3.3 mag throughout $\Delta$R.A.–$\Delta$Dec plane, even at the 2011 and 2016 positions of K2-33.

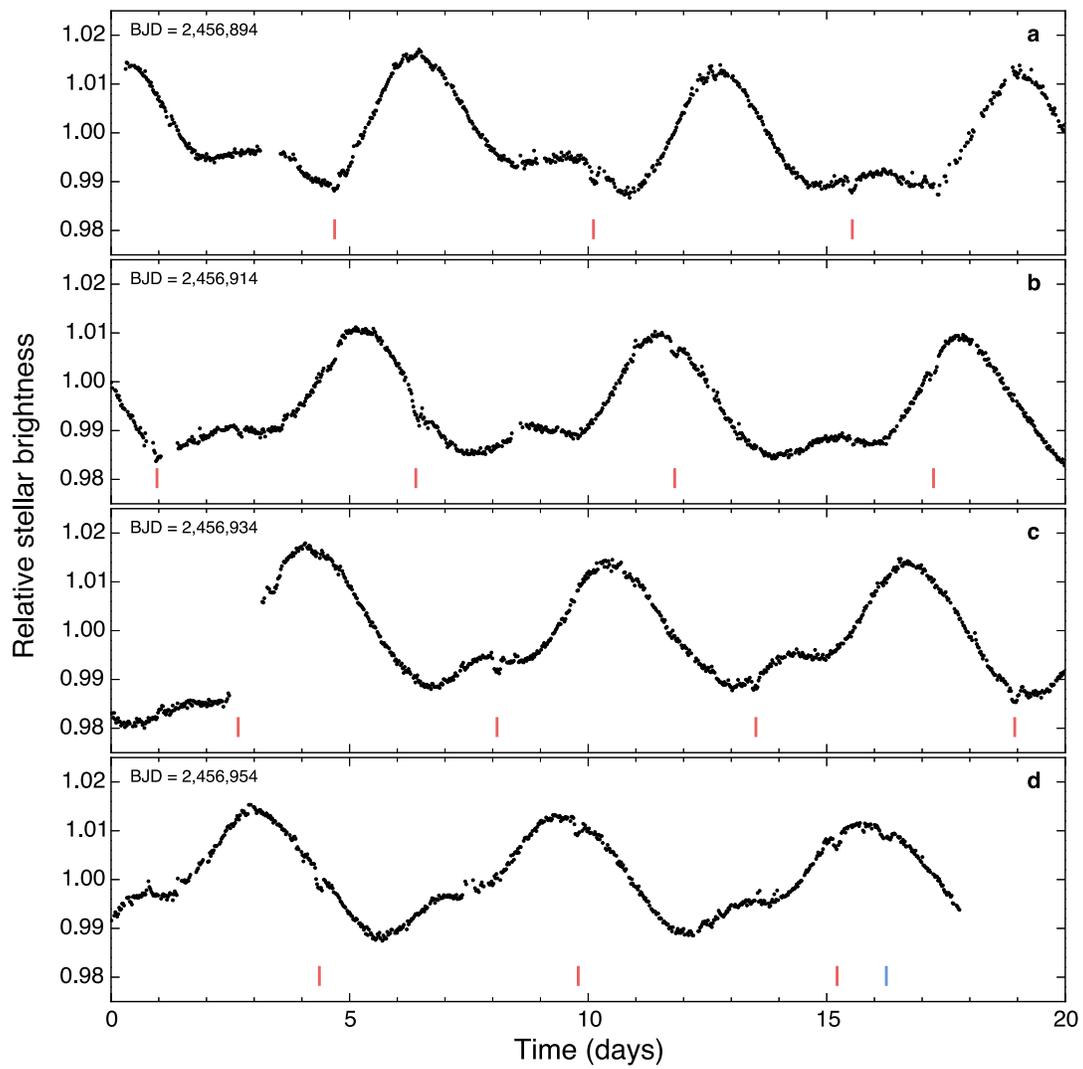
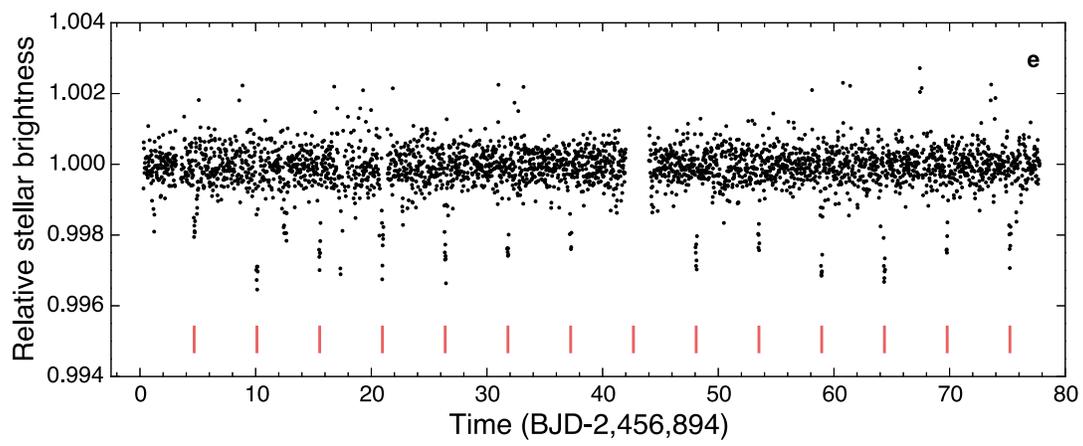
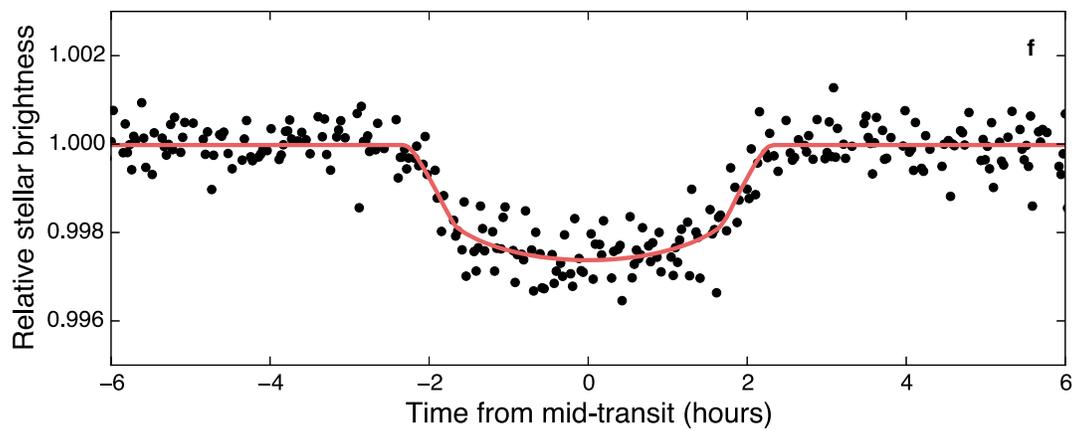

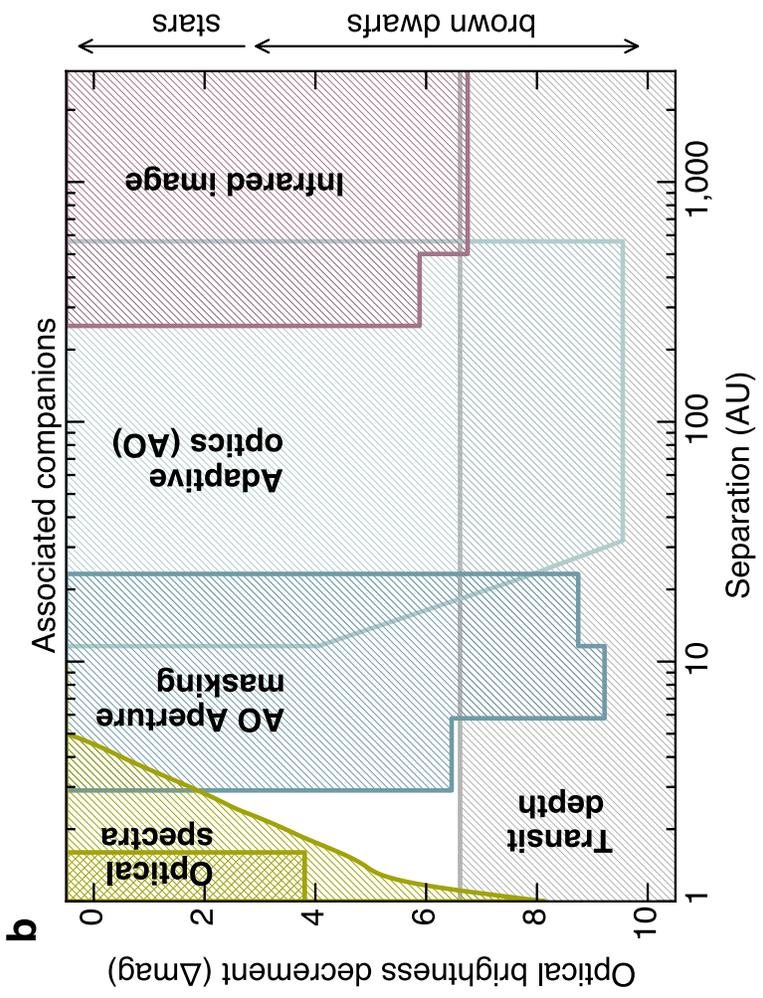
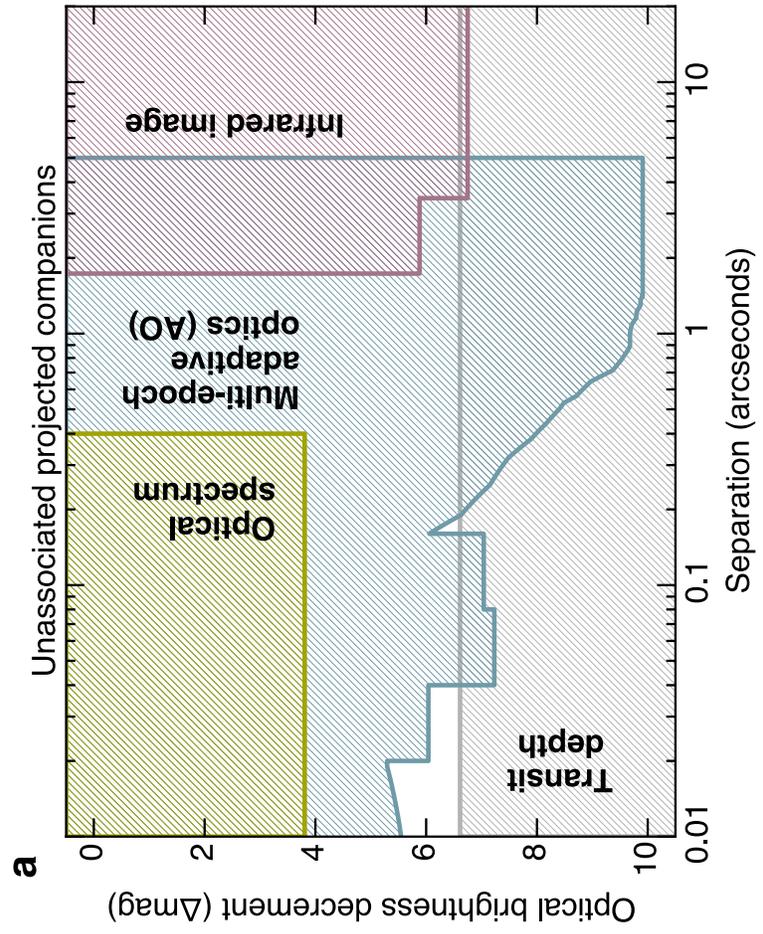

| Epoch (UTC) | BJD | Radial velocity (km s$^{-1}$) |
|---|---|---|
| 2015-06-01 12:57:36.0 | 2457175.040 | -6.61 ± 0.58 |
| 2016-02-02 15:30:14.4 | 2457421.146 | -6.66 ± 0.30 |
| 2016-02-04 15:23:02.4 | 2457423.141 | -6.60 ± 0.30 |
| 2016-02-21 14:29:45.6 | 2457440.104 | -6.40 ± 0.30 |

**Extended Data Table 1.**

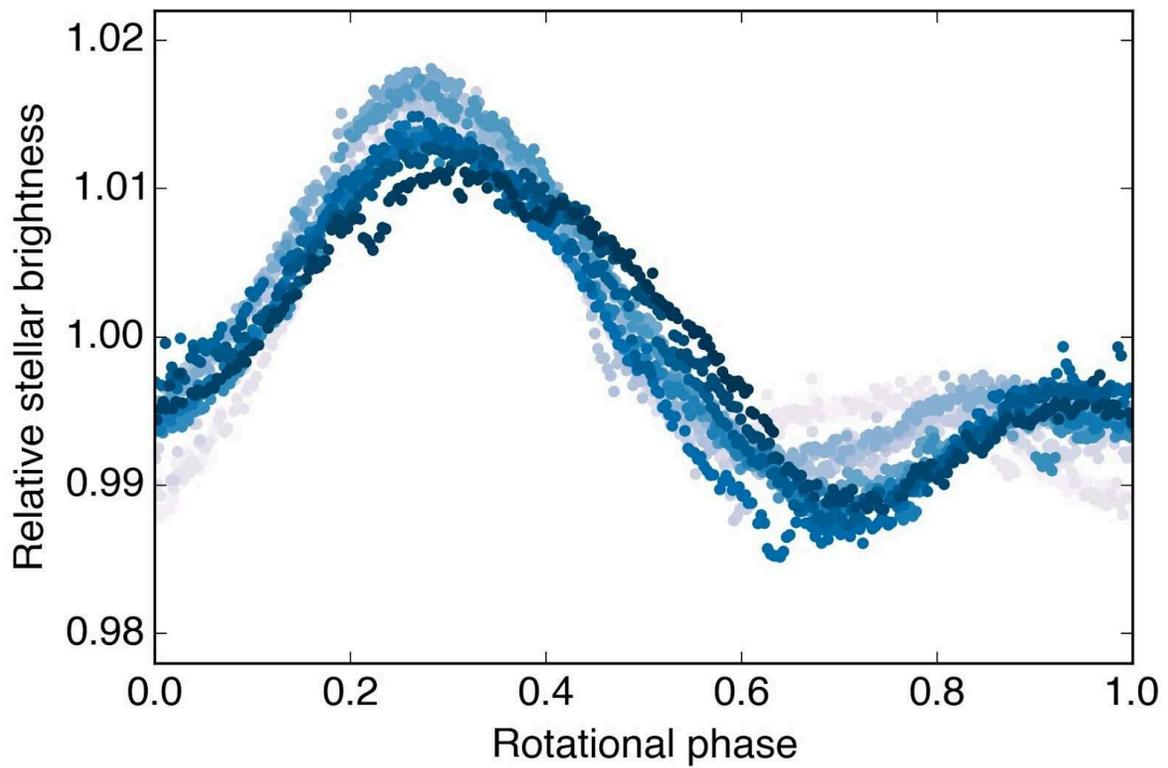

**Extended Data Figure 1.**

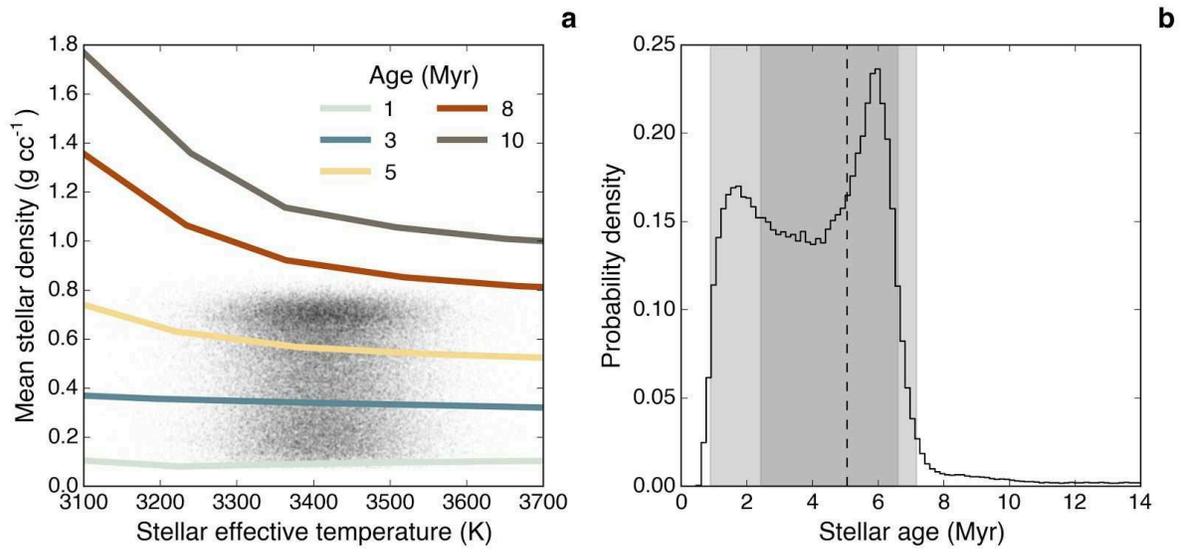

**Extended Data Figure 2.**

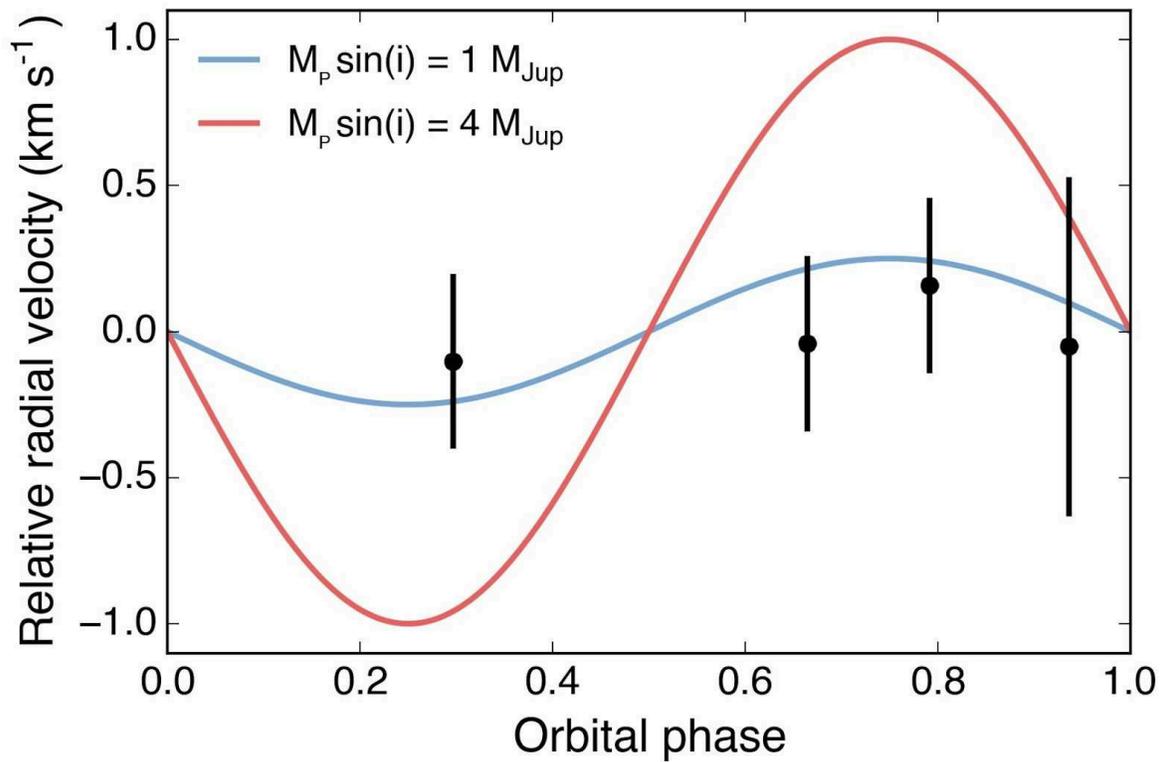

**Extended Data Figure 3**.

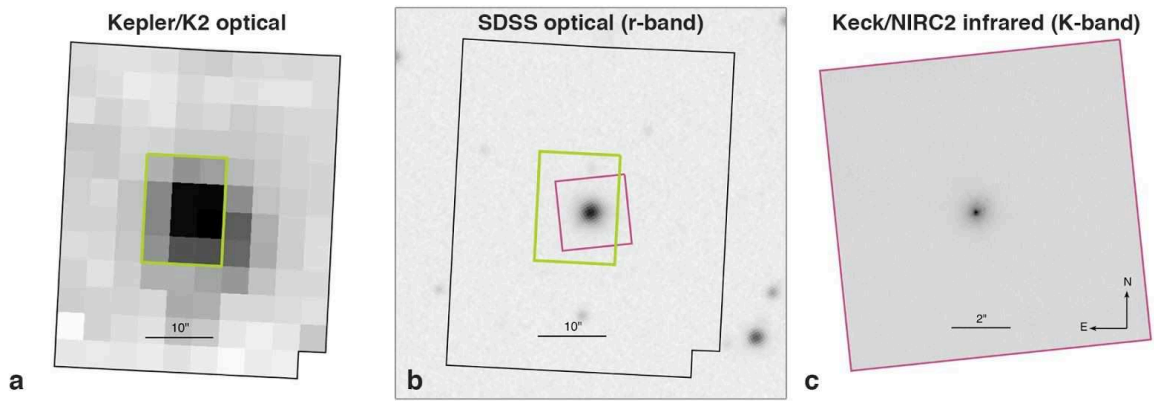

**Extended Data Figure 4.**

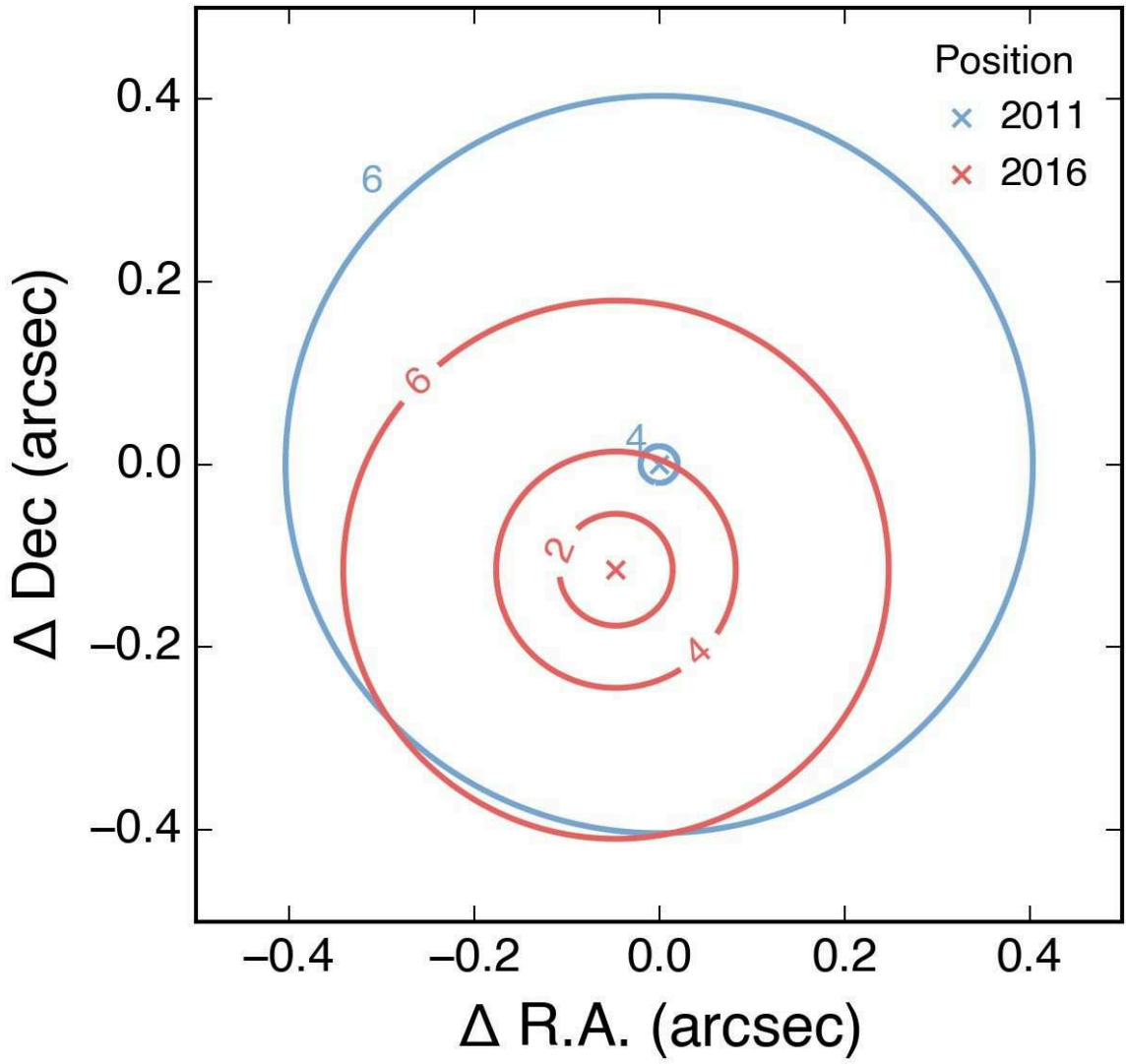

**Extended Data Figure 5.**